\documentclass[
aps,%
11pt,%
final,%
notitlepage,%
oneside,%
onecolumn,%
nobibnotes,%
nofootinbib,
superscriptaddress,%
showpacs,%
centertags]%
{revtex4}
\usepackage[cp1251]{inputenc}
\usepackage[english]{babel}
\usepackage{hhline}
\usepackage{amsmath,amssymb}
\usepackage{bm}
\usepackage{graphicx}
\begin {document}
\title{The calculation of the muon transfer rate from protium to neon\\
on the ground of a two-centre Coulomb basis}
\author{\firstname{S.~V.}~\surname{Romanov}}
\email[]{Romanov_SVi@nrcki.ru}
\email[]{Serguei.V.Romanov@gmail.com}
\affiliation{National Research Centre "Kurchatov Institute",
Moscow, 123182, Russia.}
\affiliation{Moscow Institute of Physics and Technology, Moscow,
123098, Russia.}
%
%
\begin{abstract}
The results of improved calculations of the muon transfer rate
from the $1S$--state of muonic protium to neon are presented in
the interval of collision energies from $10^{-4}$~eV to 15~eV. The
calculations have been made within the perturbed stationary states
method in which the wavefunction of the three-body system (muon,
proton and neon nucleus) is expanded in eigenfunctions of a
two-centre Coulomb problem formulated in the Jacobi coordinates of
the entrance channel. This approach provides the asymptotically
correct description of the entrance channel. Namely, the correct
dissociation limit is obtained, there are no spurious long-range
interactions, the polarization attraction between muonic protium
and neon appears naturally. Moreover, the electron screening,
which is important at low collision energies, can be easily taken
into account. The defects of the description are removed into the
muon transfer channel in which their effect is not expected to be
too significant because of large energies of the relative motion
in this channel. The previous calculations carried out in this way
allowed one to explain experimentally observed features of the
temperature dependence of the transfer rate in hydrogen-neon
mixtures. In the present work, a more perfect algorithm of
constructing the basis eigenfunctions of the two-centre Coulomb
problem has been realized and a better agreement with experimental
data has been obtained.
\end{abstract}
\pacs{
34.70.+e, 
36.10.Ee 
}
\maketitle
%
%
\section{Introduction}
\label{Intro}
%
%
The direct muon transfer from protium to neon is considered in the
present paper. It is a particular case of the charge transfer from
muonic protium $\mu p\,(1S)$ in its ground state to a chemical
element with the atomic number $Z$\,:
\begin{equation}
\label{transf}
\mu p\,(1S)+Z\to {\mu Z\,}^*+p\,.
\end{equation}
Here ${\mu Z\,}^*$ is the muonic atom of the element $Z$ in an
excited state. The rate $\lambda(T)$ of the muon transfer from
thermalized $\mu p$ atoms was measured in liquid hydrogen-neon
mixtures at the temperature $T=20$~K~\cite{Schell} and in dense
gaseous mixtures at $T=300$~K~\cite{Jacot}\,. The corresponding
results traditionally reduced to the atomic density of liquid
hydrogen $N_{\rm H}=4.25\times 10^{22}\;\mbox{\rm cm}^{-3}$ are
given in Table~\ref{tab_res}. The most interesting feature of the
%
%
%
%
\begin{table}
\caption{The experimental $\lambda_e$ and calculated $\lambda_t$
values of the rate $\lambda(T)$ of the muon transfer from
thermalized $\mu p\,(1S)$ atoms to neon for two values of the
temperature $T$. All the rates are given in units of
$10^{10}\,\mbox{\rm s}^{-1}$ and reduced to the atomic density of
liquid hydrogen. Three values of the transfer rate given in two
columns correspond to the cases A, B, and C of taking into account
the electron screening.}
\bigskip
\begin{tabular}{|@{\hspace{4mm}}c@{\hspace{4mm}}|@{\hspace{4mm}}c@{\hspace{4mm}}|c@{\hspace{4mm}}c@{\hspace{4mm}}|@{\hspace{4mm}}c@{\hspace{4mm}}|@{\hspace{4mm}}c@{\hspace{4mm}}|@{\hspace{2mm}}c@{\hspace{2mm}}|}
\hline
$T$,~K&$\lambda_e$&\multicolumn{5}{c|}{$\lambda_t$}\\
%
%
\cline{3-7}
      &\cite{Schell,Jacot}&&\cite{Sayas}&\cite{Roman}       &\cite{Fr2}&the present work\\
\hline
20    &$3.00\pm 1.00$     &&$-$         &6.40\enskip (A)    &1.2       &2.48\enskip (A)\\
      &                   &&            &1.30\enskip (B)    &          &0.81\enskip (B)\\
      &                   &&            &3.56\enskip (C)    &          &2.00\enskip (C)\\
\hline
300   &$0.849\pm 0.018$   &&0.4         &3.31\enskip (A)    &1.02      &1.62\enskip (A)\\
      &                   &&            &1.36\enskip (B)    &          &0.87\enskip (B)\\
      &                   &&            &2.02\enskip (C)    &          &1.25\enskip (C)\\
\hline
\end{tabular}
\label{tab_res}
\end{table}
%
%
reaction considered is its anomalously small rate at room
temperature. It is an order of magnitude less than the muon
transfer rate to other elements with $Z\geq 6$~\cite{Schell}\,.
Moreover, the temperature dependence of the transfer rate is of
interest. Indeed, as the temperature is increased from 20~K to
300~K, $\lambda(T)$ falls by a factor of three. This means that
the standard $v^{-1}$ dependence of the transfer cross-secton on
the relative velocity $v$ is incorrect at collision energies
corresponding to room temperature.

The first attempt to explain the anomalously small value of the
transfer rate at room temperature was made by
Sayasov~\cite{Sayas}. A traditional way of calculating the rate of
the reaction~(\ref{transf}) is based on consideration of
quasicrossings of adiabatic terms correlated to states of the
muonic atoms $\mu p\,(1S)$ and ${\mu Z\,}^*$. Then the
Landau-Zener formula is used to estimate probabilities of
transitions between these terms~\cite{Gersh}. Sayasov pointed to
the fact that, in the case of the muon transfer to light elements,
the WKB approximation is not applicable in quasicrossing regions.
Therefore, the standard Landau-Zener formula is also not
applicable. Having modified it in a proper way, Sayasov obtained
the transfer rate in the form of an oscillating function whose
argument depends on $Z$\,, the mass of the hydrogen isotope, and
the coordinate of the relevant quasicrossing point (the two-state
approximation was used and only the $S$--wave was considered). For
the muon transfer from protium to neon, the value of the argument
proved to be close to the position of a minimum of the function
and the estimated value of the transfer rate was found to be two
times less than the experimental value (Table~\ref{tab_res}).
Concerning the decrease of the transfer rate in the temperature
interval 20--300~K, it remained unexplained. The standard $v^{-1}$
law was obtained for the transfer cross-section. This means that
the transfer rate does not depend on $v$ and, consequently, it is
independent of temperature.

The next step was made in the work~\cite{Roman}. It was carried
out in connection with the experiment~\cite{Laser} on the laser
excitation of the $2S-2P$ transition in muonic protium with the
aim of a precise determination of the mean-square charge radius of
the proton. The muon transfer from the metastable $2S$--state of
muonic protium to neon was considered as a way of detecting this
state~\cite{David,Rand}. In this case, the muon transfer from the
$1S$--state is a background and it is desirable to know the energy
dependence of its rate. In ref.~\cite{Roman} this rate was
calculated in the interval of collision energies from $10^{-4}$~eV
to 15~eV. The lowest value corresponds to the temperature about
1~K, the uppermost one is close to the lowest electron excitation
energy of neon (16.6~eV~\cite{Smirn}). The method of calculations
was based on the substantial difference in energies of the
relative motion in reaction channels. In the entrance channel
$\bigl(\mu p\,(1S)+{\rm Ne}\bigr)$ the collision energy does not
exceed 15~eV, whereas in the transfer channel $\bigl({\mu{\rm
Ne}\,}^*+p\bigr)$ it is a~few~keV~\cite{Jacot}\,. It is obvious
that an asymptotically correct description of the entrance channel
is of primary importance in this case. Accordingly, the wave
function of the three-body system was constructed as an expansion
in eigenfunctions of a two-centre Coulomb problem formulated in
the Jacobi coordinates of the entrance channel. As a result, the
correct dissociation limit is obtained in this channel, no
spurious long-range interactions arise, and the polarization
attraction between muonic protium and neon appears naturally.
Moreover, even in the simplest approximation the dipolar
polarizability of muonic protium is reproduced with one percent
accuracy. It is also significant that the electron screening in
the entrance channel can be easily taken into account. It proves
to be important at low collision energies.

A disadvantage of this approach is that the transfer channel is
described in unnatural coordinates (in the Jacobi coordinates of
the entrance channel). The eigenstates of the two-centre problem
localized at the neon nucleus in the separated atoms limit are not
eigenstates of the Hamiltonian of isolated muonic neon.
Nevertheless, the inclusion of a group of such states in
calculations allows one to describe the migration of the muon from
protium to neon. It is obvious that no partial transfer rates to
individual states of the final ${\mu\rm Ne\,}^*$ atom can be
obtained in this way. However, the total transfer rate can be
evaluated. Although the asymptotic description of the transfer
channel shows a number of defects (incorrect dissociation limits,
spurious long-range interactions), their effect is not expected to
be too significant because of large energies of the relative
motion in this channel. In truth, the method employed in
ref.~\cite{Roman} is a variant of the well-known perturbed
stationary states (PSS) method. However, unlike its standard
realization~\cite{Ponom1} in which all the binary channels are
described incorrectly in asymptotic domains, the above approach
provides the asymptotically correct description of the entrance
channel with low collision energies and removes all the
difficulties into the muon transfer channel.

The calculation made in ref.~\cite{Roman} with four basis
eigenfunctions of the two-centre problem recognized some features
of the muon transfer from protium to neon.
\begin{enumerate}
\item The transfer rate treated as a function of the collision
energy has a well pronounced minimum at thermal energies
($T=300$~K). This corresponds to the above-mentioned strong
suppression of the transfer reaction at room temperature.
\item At the same energies the contribution of the $P$--wave to
the transfer rate becomes significant (20--30~\%). At the
subsequent energy growth, the contributions of waves with greater
angular momenta increase rapidly. This leads to the transfer rate
going up at energies greater than 0.1~eV. In particular, a
resonance peak appears at collision energies of 0.3--0.5~eV. It is
due to the existence of a quasi-steady state in the $D$--wave. It
should be noted that only the $S$--wave was considered in most of
earlier calculations.
\item The electron screening in the entrance channel proves to be
important at the collision energies less than 1~eV. In order to
clarify its role, the calculations were made for the following
three cases.
\begin{enumerate}
\item[A)] The electron screening was fully ignored. This
corresponds to the muon transfer to a bare neon nucleus. The
interaction of muonic protium with neon at large separations was
described with the help of the ordinary potential of the
polarization attraction.
\item[B)] The screening of the nuclear charge of neon by atomic
electrons was taken into account in the polarization potential.
\item[C)] A contact interaction of muonic protium with the
electron shell of neon was added to the screened polarization
potential. This interaction is due to the finite size of muonic
protium, and it is proportional to the product of the mean-square
charge radius of $\mu p\,(1S)$ and the electron density of neon.
It leads to an additional attraction. This case is most realistic
because the electron screening is taken into account in a maximum
degree.
\end{enumerate}
\end{enumerate}
After averaging over the Maxwellian distribution, the rate
$\lambda(T)$ of the muon transfer from thermalized $\mu p$ atoms
was obtained. Its values for the temperatures of 20~K and 300~K
are given in Table~\ref{tab_res}. Attention should be paid to the
strong dependence of the results on the way of considering the
electron screening. In passing from the case A to the case B, the
attraction in the entrance channel becomes weaker and the transfer
rate decreases. This is most noticeable at low temperatures. The
additional attraction in the case C leads to the transfer rate
increasing. Concerning the agreement with experimental data, it is
good at $T=20$~K in the most realistic case C. Of course, it is
necessary to note that in a liquid hydrogen-neon mixture the
electron screening may be more complicated than it was assumed in
the calculation. Moreover, the Maxwellian distribution seems to be
a too crude model in this case. At $T=300$~K the agreement is
worse: the transfer rate calculated in the case C exceeds the
experimental value by a factor of 2.3\,. Nevertheless, the
calculation correctly reproduces the tendency to decreasing the
transfer rate with increasing the temperature. It is interesting
that the transfer rate calculated in the case B is nearly constant
in the interval 20--300~K, although the agreement with the
experimental data at room temperature is better.

After the paper~\cite{Roman} had come out, the results of
calculations made within a hyperspherical elliptic coordinates
method were published~\cite{Fr1,Fr2}. The values of the muon
transfer rate from protium to neon obtained in these works are
also given in Table~\ref{tab_res}. Having included more than one
hundred basis functions in the calculation, the authors obtained a
very good result. Namely, the observed value of the transfer rate
at room temperature was reproduced with the accuracy of 15~\%. The
agreement with the experimental data at $T=20$~K is worse: the
calculated value of the transfer rate is less than a half of the
observed value. Moreover, the transfer rate is nearly constant in
the temperature interval considered. It is important to note that
the electron screening was fully ignored in this calculation, i.e.
the muon transfer to the bare neon nucleus was considered.
Introducing the screening may noticeably reduce calculated values
of the transfer rate. For example, according to ref.~\cite{Ital},
at room temperature the electron screening reduces the transfer
rate by a factor of $Z^{\,1/3}$, i.e. nearly in half for neon.
Such a reduction has been demonstrated in ref.~\cite{Roman} in
passing from the case A to the case B.

This paper is a sequel to the work~\cite{Roman}. It presents the
results of calculations made within the same approximations but
improved in one point. Namely, a more perfect algorithm of
constructing the basis eigenfunctions of the two-centre Coulomb
problem has been realized.
%
%
\section{Description of the method}
\label{Meth}
%
%
As the calculation method has been detailed in ref.~\cite{Roman},
only its main points will be briefly considered here. Unless
otherwise specified, muon-atom units (m.a.u. for short) are used
below:
\begin{equation}
\label{mau}
\hbar=e=m_\mu=1\,,
\end{equation}
$e$ is the proton charge, $m_\mu$ is the muon mass; the length and
energy units are respectively equal to $2.56\times10^{-11}$~cm and
5.63~keV.

\begin{figure}[]
\includegraphics{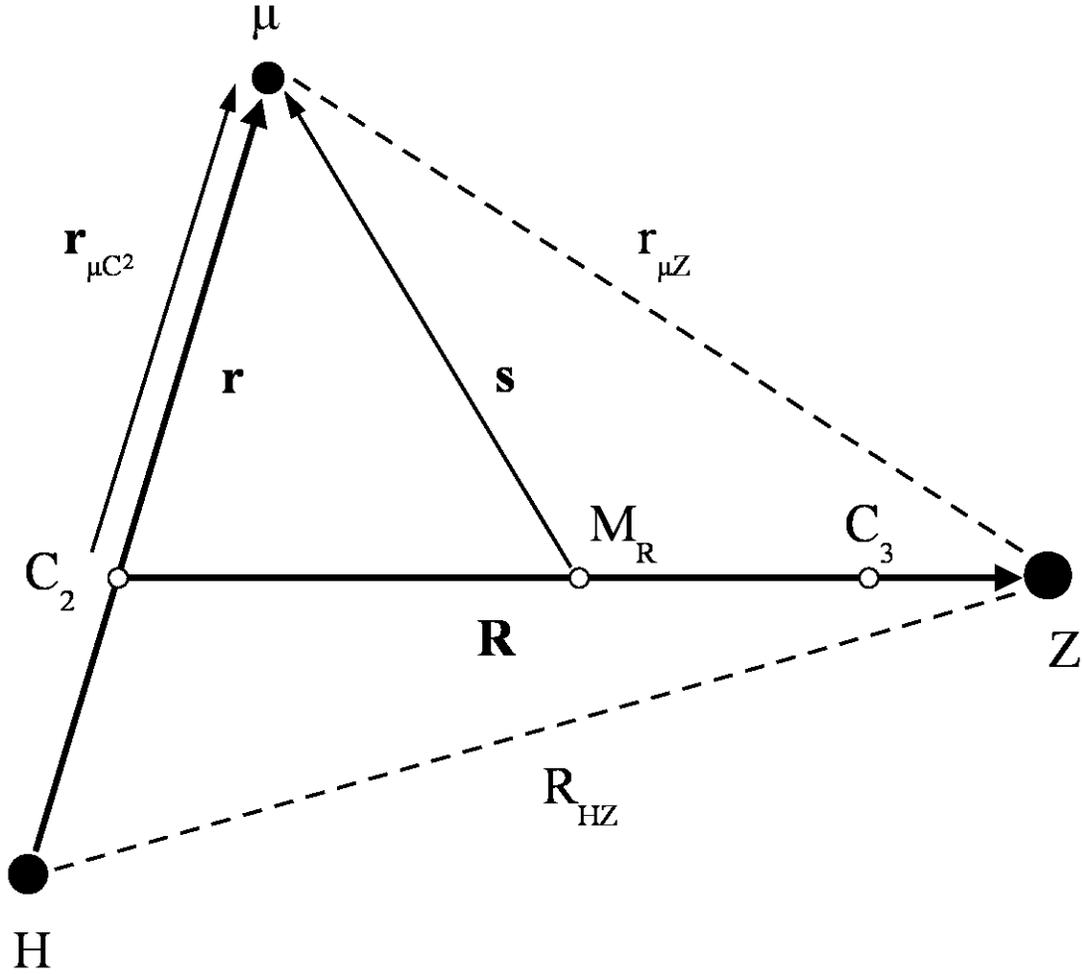}
\caption{The Jacobi coordinates of the entrance channel and other
notations. $C_2$ is the centre of mass of muonic hydrogen, $C_3$
is the centre of mass of the three-body system, $M_R$ is the
midpoint of the vector $\bf R$.}
\label{notations}
\end{figure}

Let us consider the system consisting of a negative muon $\mu$, a
nucleus $\rm H$ of a hydrogen isotope, and a nucleus with the
atomic number $Z$. Let us introduce the Jacobi coordinates of the
entrance channel of the reaction~(\ref{transf}): the vector $\bf
r$ connecting the nucleus $\rm H$ with the muon and the vector
$\bf R$ joining the centre of mass $C_2$ of muonic hydrogen
$\mu\rm H$ and the nucleus $Z$ (Fig.~\ref{notations}). The centre
of mass
%
%
%
%
%
%
$C_3$ of the three-body system lies on the vector $\bf R$. In the
centre-of-mass frame the Hamiltonian of this system is:
\begin{equation}
\label{Htot}
\hat H=-\frac{1}{2M_r}\,{\Delta}_{\bf R}+{\hat H}_\mu
+\frac{Z}{R_{{\rm H}Z}}\,.
\end{equation}
The first term is the operator of the kinetic energy of the
relative motion of muonic hydrogen and the nucleus $Z$. $M_r$ is
the reduced mass of the nucleus $Z$ with respect to muonic
hydrogen:
\begin{equation}
\label{Mr}
M_r^{-1}=(M_{\rm H}+1)^{-1}+M_Z^{-1}\,,
\end{equation}
$M_{\rm H}$ and $M_Z$ are the nuclear masses. ${\hat H}_\mu$ is
the Hamiltonian of muonic hydrogen with the addition of the
Coulomb attraction of the muon and the nucleus $Z$:
\begin{equation}
\label{Hmu1}
{\hat H}_\mu=-\frac{1}{2m_{\mu\rm H}}\,{\Delta}_{\bf r}
-\frac{1}{r}-\frac{Z}{r_{\mu Z}}\,,
\end{equation}
$m_{\mu\rm H}$ is the reduced mass of muonic hydrogen:
\begin{equation}
\label{mrH}
m_{\mu\rm H}^{-1}=M_{\rm H}^{-1}+1\,,
\end{equation}
$r_{\mu Z}$ is the distance between the muon and the nucleus $Z$.
The last term in the formula~(\ref{Htot}) is the Coulomb repulsion
of the nuclei $\rm H$ and $Z$, $R_{{\rm H}Z}$ is the internuclear
distance.

Let us isolate a two-centre problem in the three-body Hamiltonian.
For this purpose the term ${\hat H}_\mu$ is rewritten as
follows~\cite{Jap}:
\begin{equation}
\label{Hmu2}
{\hat H}_\mu=m_{\mu\rm H}\cdot{\hat h}_\mu\,,
\end{equation}
\begin{equation}
\label{Hmu3}
{\hat h}_\mu=-\frac{1}{2}\,{\Delta}_{\bf s} -\frac{1}{|{\bf
s}+\frac{\bf R}{2}|} -\frac{Z\,'}{|{\bf s}-\frac{\bf R}{2}|}\,.
\end{equation}
The vector $\bf s$ connects the midpoint $M_R$ of the vector $\bf
R$ with the muon (Fig.~\ref{notations}):
\begin{equation}
\label{s}
{\bf s}={\bf r}_{\mu C_2}-\frac{\bf R}{2}\,,
\quad
{\bf r}_{\mu C_2}=m_{\mu\rm H}\cdot{\bf r}\,.
\end{equation}
The vector ${\bf r}_{\mu C_2}$ joins the centre of mass of muonic
hydrogen and the muon. The quantity $Z\,'$ is:
\begin{equation}
\label{Z'def}
Z\,'=\frac{Z}{m_{\mu\rm H}}\,.
\end{equation}
${\hat h}_\mu$ is the Hamiltonian of the muon in the field of two
Coulomb centres whose charges are equal to unity and $Z\,'$. The
unit charge is placed in the centre of mass of muonic hydrogen,
the position of the charge $Z\,'$ coincides with the one of the
nucleus $Z$. For the muon transfer from protium to neon
\begin{equation}
\label{Z'neon}
m_{\mu\rm H}\approx 0.899\,,
\quad
Z\,'\approx 11.1\,.
\end{equation}
In the coordinate frame with the origin in the point $M_R$ and the
polar axis directed along the vector $\bf R$\,, the position of
the muon is specified by prolate spheroidal coordinates $\xi$,
$\eta$ and $\varphi$~\cite{Ponom2}:
\begin{equation}
\label{psc}
\xi=\frac{r_{\mu C_2}+r_{\mu Z}}{R}\,,
\quad
\eta=\frac{r_{\mu C_2}-r_{\mu Z}}{R}\,,
\end{equation}
$R$ is the length of the vector $\bf R$. The azimuthal angle
$\varphi$ lies in the plane passing through the point $M_R$
perpendicularly to $\bf R$. Surfaces of constant values of the
coordinates $\xi$ and $\eta$ are prolate ellipcoids of revolution
and two-sheeted hyperboloids with the focuses in the points $C_2$
and $Z$. For these points $\xi=1$ and $\eta=\mp 1$.

Let us consider the eigenvalue and eigenfunction problem for the
two-centre Hamiltonian ${\hat h}_\mu$:
\begin{equation}
\label{2cent} {\hat h}_\mu\,\psi_{jm}(\xi,\eta;R)\,\frac{\exp (\pm
im\varphi)} {\sqrt{2\pi}}={\varepsilon_{jm}}(R)\,
\psi_{jm}(\xi,\eta;R)\,\frac{\exp (\pm im\varphi)}{\sqrt{2\pi}}
\,.
\end{equation}
The dependence on the angle $\varphi$ is explicitly indicated
here, $m$ is a nonnegative integer, the subscript $j$ denotes a
set of the other quantum numbers. For bound states these are
either the numbers $n_\xi$ and $n_\eta$ of nodes in the
corresponding variables or the parabolic quantum numbers $n_1$ and
$n_2$ in the limit $R\to\infty$~\cite{Ponom2}. The two-centre
problem~(\ref{2cent}) is solved at a fixed distance $R$ which
appears in eigenfunctions and eigenvalues as a parameter. The
eigenfunctions with the same $m$ and different sets $i$ and $j$ of
the other quantum numbers are orthonormal:
\begin{gather}
\label{orto}
\int\psi_{im}(\xi,\eta;R)\,\psi_{jm}(\xi,\eta;R)\,d\tau=
\delta_{ij}\,;\notag\\
d\tau=(R/2)^3\,({\xi}^2-{\eta}^2)\,d\xi\,d\eta\,.
\end{gather}
The integral is taken over the region $1\le\xi<\infty$\,,
$-1\le\eta\le+1$\,. The orthonormalization with respect to $m$ is
provided by the factors $\exp(\pm im\varphi)/\sqrt{2\pi}$. It is
obvious that the solutions of the problem~(\ref{2cent}) are the
eigenfunctions of the Hamiltonian ${\hat H}_\mu$ with the
eigenvalues \mbox{$m_{\mu\rm H}\cdot\varepsilon_{jm}(R)$}.

It is well known that the two-centre problem permits separation of
variables in the prolate spheroidal coordinates~\cite{Ponom2}.
Every eigenfunction $\psi_{jm}(\xi,\eta;R)$ is the product of
radial and angular functions depending separately on $\xi$ and
$\eta$. Solving a pair of differential equations for these
functions under suitable boundary conditions allows one to find
the eigenvalue ${\varepsilon_{jm}}(R)$ and the separation constant
and, finally, to construct the functions. In ref.~\cite{Roman}
this procedure was realized on the basis of comparison equations
suggested in ref.~\cite{Solov}. The method of solving the
two-centre problem was improved in the present work. Namely, an
algorithm based on the well--known infinite
expansions~\cite{Jaffe,Baber,Bates} of the radial and angular
functions was implemented in practice. The eigenvalue
${\varepsilon_{jm}}(R)$ and the separation constant were
determined according to the method suggested in ref.~\cite{Had}
and modified for the case in which expansion coefficients are
nonmonotonic functions of their number.

The three-body system is considered in an initial coordinate frame
with fixed axes and the origin in the centre of mass $C_3$\,. Let
us introduce the operator $\hat{\bf J}$ of the orbital angular
momentum of the three-body system. The Hamiltonian $\hat H$
commutes with the operator ${\hat{\bf J}}^2$ of its square and
with the operator ${\hat J}_z$ of its projection on the $z$--axis
of the initial frame. Moreover, $\hat H$ commutes with the
operator $\hat{\rm P}$ of the coordinate inversion. A convenient
basis in which the three-body wavefunction is expanded consists of
eigenfunctions of these three operators. Let us require them to be
eigenfunctions of the two-centre problem~(\ref{2cent}). As the
spheroidal coordinates of the muon are defined with respect to the
vector $\bf R$\,, let us introduce the polar angle $\Theta$ and
the azimuthal angle $\Phi$ specifying the direction of $\bf R$ in
the initial coordinate frame. Then a configuration of the
three-body system is specified by the independent variables $R$,
$\Theta$, $\Phi$, $\xi$, $\eta$, $\varphi$\,; and the basis
functions are:
\begin{equation}
\label{basis}
\Psi_{Mjm}^{JP}(R,\Theta,\Phi,\xi,\eta,\varphi)=
\frac{{\chi}_{jm}^{JP}(R)}{R}\,\Upsilon_{Mm}^{JP}(\Phi,\Theta,
\varphi)\,\psi_{jm}(\xi,\eta;R)\,.
\end{equation}
${\chi}_{jm}^{JP}(R)$ is a radial function depending on the
indicated quantum numbers,
$\Upsilon_{Mm}^{JP}(\Phi,\Theta,\varphi)$ is the eigenfunction of
the operators ${\hat{\bf J}}^2$, ${\hat J}_z$\,, and $\hat{\rm P}$
with the eigenvalues $J(J+1)$, $M$, and $P$. The nonnegative
integer $m$ introduced in~(\ref{2cent}) is the modulus of the
projection of the angular momentum on the direction of the vector
$\bf R$. The functions $\Upsilon_{Mm}^{JP}$ are orthonormal:
\begin{equation}
\label{ornY}
\int\limits_0^{\pi}\sin\Theta\,d\Theta
\int\limits_0^{2\pi}d\Phi\int\limits_0^{2\pi}d\varphi\,
\bigl(\Upsilon_{Mm}^{JP}\bigr)^*\,\Upsilon_{M'm'}^{J'P'}=
\delta_{JJ'}\,\delta_{PP'}\,\delta_{MM'}\,\delta_{mm'}\,.
\end{equation}
Their form depends on $m$. If $m=0$, then
\begin{equation}
\label{Y_M0}
\Upsilon_{Mm=0}^{JP}(\Phi,\Theta,\varphi)=
\frac{Y_{JM}(\Theta,\Phi)}{\sqrt{2\pi}}\,,
\end{equation}
$Y_{JM}(\Theta,\Phi)$ is the ordinary spherical function. In this
case the parity is unambiguously specified by the quantum number
$J$: $P=(-1)^J$. If $m\neq 0$, then
\begin{equation}
\label{Y_Mm}
\Upsilon_{Mm}^{JP}(\Phi,\Theta,\varphi)=
\frac{\sqrt{2J+1}}{4\pi}\left[
(-1)^mD_{Mm}^J(\Phi,\Theta,\varphi)+
P(-1)^JD_{M(-m)}^J(\Phi,\Theta,\varphi)\right],
\end{equation}
$D_{Mm}^J$ and $D_{M(-m)}^J$ are the Wigner functions~\cite{Dav}
transformed under the inversion as follows:
\begin{equation}
\label{InvD} D_{Mm}^J(\Phi,\Theta,\varphi)\longrightarrow
(-1)^{J-m}D_{M(-m)}^J(\Phi,\Theta,\varphi)\,.
\end{equation}
In this case the two values of the parity are possible at given
$J$: $P=\pm (-1)^J$.

Let us consider the time-independent Schr{\"o}dinger equation for
the three-body wavefunction with the quantum numbers $J$, $M$, and
$P$:
\begin{equation}
\label{Schr}
\hat H\,\Psi_{M}^{JP}=E\,\Psi_{M}^{JP}\,.
\end{equation}
For the reaction~(\ref{transf}) the energy is:
\begin{equation}
\label{Etot}
E=E_{\mu\rm H}(1S)+E_c\,.
\end{equation}
$E_{\mu\rm H}(1S)$ is the energy of the ground state of muonic
hydrogen:
\begin{equation}
\label{EmuH}
E_{\mu\rm H}(1S)=-\frac{m_{\mu\rm H}}{2}\,,
\end{equation}
$E_c$ is the collision energy:
\begin{equation}
\label{Ec}
E_c=\frac{M_r v^2}{2}=\frac{k^2}{2M_r}\,,
\end{equation}
$v$ is the velocity of the relative motion of the $\mu\rm H$ atom
and the nucleus $Z$ at infinite separation, $k={M_r}v$ is the
asymptotic momentum of the relative motion.

Let us seek a solution of the Schr{\"o}dinger equation in the form
of an expansion in the basis functions~(\ref{basis}):
\begin{equation}
\label{expan}
\Psi_M^{JP}=\sum_{jm}\,\Psi_{Mjm}^{JP}\,.
\end{equation}
The substitution of this expansion into the equation~(\ref{Schr})
and the integration over the variables $\Theta$, $\Phi$, $\xi$,
$\eta$, and $\varphi$ under the orthonormalization
condition~(\ref{ornY}) yield a set of coupled second-order
differential equations for the radial functions
${\chi}_{jm}^{JP}(R)$. These equations are given in
ref.~\cite{Roman}. In practice, a finite number of two-centre
states is taken into account in the expansion~(\ref{expan}).
Solving the obtained set of coupled equations under suitable
boundary conditions allows one to calculate the total
cross-section of the reaction~(\ref{transf}).

As already noted, the main idea of the present approach is to
provide the asymptotically correct description of the entrance
channel of the muon transfer reaction at large distances $R$. In
the limit~$R\to\infty$ the bound eigenstates of the two-center
problem~(\ref{2cent}) fall into two groups. The states of one
group are localized near the left centre, which is placed in the
centre-of-mass of the $\mu\rm H$ atom and has the unit charge. The
states of another group are localized near the right centre
$Z\,'$. The simplest way to describe the entrance channel is to
take into account the only state of the left-centre group. Its
asymptotic quantum numbers are:
\begin{equation}
\label{1sH}
m=n_1=n_2=0\,,\quad n=1\,.
\end{equation}
$n_1$ and $n_2$ are the parabolic quantum numbers~\cite{Ponom2},
$n=n_1+n_2+m+1$ is the principle quantum number. All the
quantities related to this state will be marked with the subscript
0. In the limit considered, the eigenfunction $\psi_0$ and the
eigenvalue $\varepsilon_0(R)$ of the two-centre problem are:
\begin{equation}
\label{psi0}
\psi_0\propto\exp\,(-m_{\mu\rm H}\cdot r)\,,\quad
\varepsilon_0(R\to\infty)=-\frac{1}{2}\,.
\end{equation}
Thus, the two-centre eigenfunction goes into the wavefunction of
the ground state of muonic hydrogen with the correct reduced mass.
This is due to the left centre being placed in the centre-of-mass
of muonic hydrogen. The argument of the exponent in the function
$\psi_0$ is the distance from this centre to the muon. The
eigenvalue of the Hamiltonian ${\hat H}_\mu$ tends to the correct
dissociation limit:
\begin{equation}
\label{dis0}
m_{\mu\rm H}\cdot\varepsilon_0(R\to\infty)=E_{\mu\rm H}(1S)\,.
\end{equation}

At large $R$ the relative motion in the entrance channel is
governed by the potential $U_0(R)$ which is a result of averaging
the three-body Hamiltonian over the state $\psi_0$. The expansion
of this potential in powers of $R^{-1}$ was considred in
ref.~\cite{Roman}. Its leading term is proportional to $R^{-4}$
and corresponds to the polarization attraction of muonic hydrogen
and the nucleus $Z$:
\begin{equation}
\label{U0}
U_0(R)=-\frac{\beta_{0} Z^2}{2R^4}\,.
\end{equation}
The following value was obtained for the dipolar polarizability of
muonic hydrogen:
\begin{equation}
\label{beta0}
\beta_0=\beta\left[1-\frac{1}{(M_{\rm H}+1)^2}\right]\,,
\end{equation}
$\beta$ is the exact value of the polarizability:
\begin{equation}
\label{beta1}
\beta=\frac{9}{2m^3_{\mu\rm H}}\,.
\end{equation}
It should be noted that, because of the cube of the reduced mass
in the denominator of this formula, the value of $\beta$ may
differ noticeably from the frequently used value of 4.5\,, which
corresponds to an infinitely heavy nucleus $\rm H$. In particular,
$\beta\approx 6.20$ for muonic protium. Although $\beta_0$ is not
equal to $\beta$, their values are very close. For muonic protium
$\beta_0\approx 0.99\,\beta$\,. The difference of these values is
due to the Coulomb repulsion of the nuclei being nondiagonal in
the two-centre basis. It was shown in ref.~\cite{Roman} that
taking into account this fact yielded a small correction whose
addition to $\beta_0$ faithfully reproduces the polarizability
$\beta$. Thus, the use of the only left-centre state already
provides a good description of the entrance channel at large $R$:
the dissociation limit is correct, no spurious long-range
interactions appear (at least in the terms up to $R^{-4}$
inclusive), the polarizability of muonic hydrogen is reproduced
with one percent accuracy. Therefore, this description will be
followed below. In addition, as the values of $\beta$ and
$\beta_0$ agree closely with each other, the polarization
potential with the exact value of $\beta$ will be used to describe
the relative motion in the entrance channel at large $R$:
\begin{equation}
\label{Up}
U_p(R)=-\frac{\beta Z^2}{2R^4}\,.
\end{equation}

In the approach considered, the muon transfer channel is described
by right-centre states. In the limit $R\to\infty$ they correspond
to the $\mu Z\,'$ atom with an infinitely heavy nucleus, but not
to the real $\mu Z$ atom. In particular, the wavefunctions of
these states do not include the reduced mass of the $\mu Z$ atom
at all. Moreover, the equations for the radial functions of the
transfer channel remain coupled at $R\to\infty$\,, although, as it
was found in ref.~\cite{Roman}, coupling matrix elements are not
too large compared to the energy released in the transfer
reaction. The reason of these difficulties lies in the transfer
channel being described in the Jacobi coordinates of the entrance
channel. It is obvious that no partial transfer cross-sections to
individual states of the $\mu Z$ atom can be calculated in this
case. Nevertheless, as at large $R$ the right-centre states are
localized near the nucleus $Z$, whose position coincides with the
one of the charge $Z\,'$, a group of these states as a whole
describes the migration of the muon charge cloud from $\rm H$ to
$Z$, i.e. the muon transfer. Therefore, it is possible to
calculate the total transfer cross-section. Let us consider how
this is done.

The matrix elements of the three-body Hamiltonian coupling the
entrance and transfer channels fall exponentially at large $R$.
Therefore, in the limit $R\to\infty$ the set of radial equations
is split into two blocks which correspond to the entrance and
transfer channels. In the simplest approximation in which the only
left-centre state with the quantum numbers~(\ref{1sH}) is taken
into account, the entrance channel is described by one equation:
\begin{equation}
\label{xi0}
\frac{d^2{\chi}_0^J}{dR^2}+ \left[\,k^2-\frac{J(J+1)}{R^2}
-2M_r\,U_p(R)\right]\chi_0^J=0\,.
\end{equation}
${\chi}_0^J$ is the radial function of the entrance channel. Its
superscript $P$ is omitted because the parity is now specified by
the quantum number $J$: $P=(-1)^J$. The boundary condition at
large $R$ is:
\begin{equation}
\label{xi0as}
{\chi}_0^J\,(R\to\infty)\longrightarrow\sin(kR-J\pi/2)+Q_0^J\exp
\left[\,i(kR-J\pi/2)\,\right]\,.
\end{equation}
The complex amplitude $Q_0^J$ depends on $J$ and $k$. The radial
functions of the transfer channel are asymptotically represented
by outgoing scattered waves. A method of constructing such
solutions for coupled equations was described in
ref.~\cite{Roman}. The boundary condition at $R=0$ is standard:
all the radial functions are equal to zero in this point.

The integration of coupled equations under the above boundary
conditions allows one to construct the amplitudes $Q_0^J$. Let us
rewrite the asymptotic radial function ${\chi}_0^J$ in the
following form:
\begin{equation}
\label{xi0as1}
{\chi}_0^J(R\to\infty)\longrightarrow\frac{i}{2}\left[
\exp\left(-kR+\frac{J\pi}{2}\right)-S_0^J\exp\left(kR-\frac{J\pi}{2}
\right)\right]\,.
\end{equation}
$S_0^J$ is the diagonal $S$--matrix element corresponding to the
entrance channel:
\begin{equation}
\label{S0J}
S_0^J=1+2i\,Q_0^J\,.
\end{equation}
As the muon transfer is the only inelastic channel at the
collision energies considered, the total transfer cross-section
is~\cite{Dav}:
\begin{equation}
\label{crsec}
\sigma(E_c)=\frac{\pi}{k^2}\sum_{J=0}^{\infty}(2J+1)
\left(1-{|S_0^J|}^2\right)\,.
\end{equation}
The muon transfer rate treated as a function of the collision
energy and reduced to the atomic density of liquid hydrogen is:
\begin{equation}
\label{rtE}
q(E_c)=N_{\rm H}\,v\,\sigma(E_c)\,.
\end{equation}
The transfer rate $\lambda(T)$ from thermalized $\mu\rm H$ atoms
is obtained by averaging this quantity over the Maxwellian
distribution of relative velocities in the entrance channel.
%
%
\section{Some details of the calculations}
\label{Det}
\begin{figure}
\includegraphics{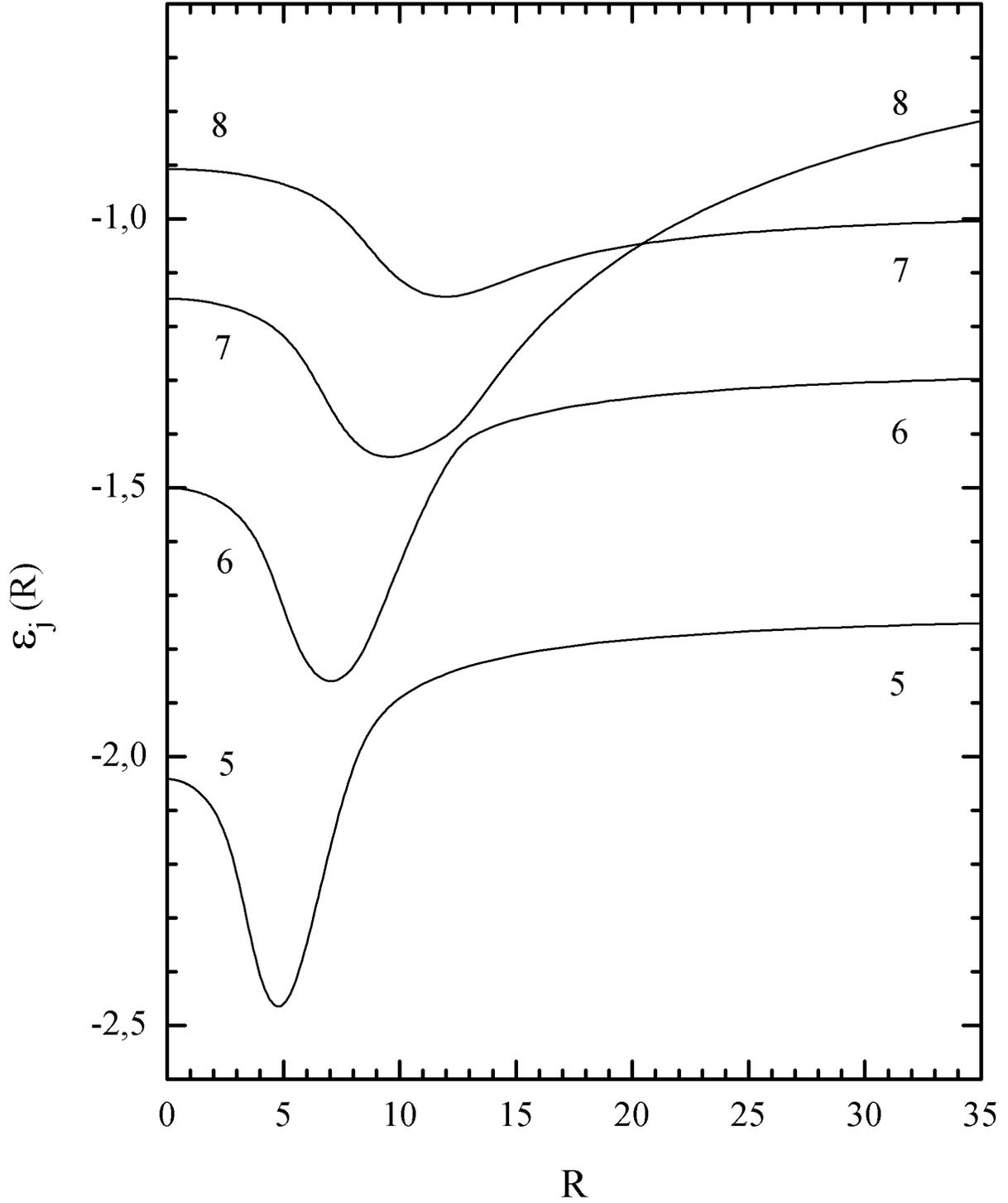}
\caption{The eigenvalues ${\varepsilon_j}(R)$ of the two-centre
Coulomb problem vs. the interatomic distance $R$ for the
right-centre states with the parabolic quantum numbers $m=n'_1=0$
and $n'_2=5-8$. All the quantities are given in m.a.u. The curves
are marked with values of the number $n'_2$.}
\label{eps_R}
\end{figure}

%
%
Let us consider how the method described in Section~\ref{Meth} is
applied to the calculation of the rate of the muon transfer from
protium to neon. At large interatomic separations, the entrance
channel is described by the only left-centre state $\psi_0$ with
the quantum numbers~(\ref{1sH}). In order to choose the relevant
right-centre states, let us take advantage of the standard
viewpoint that the muon transfer from hydrogen to a heavier
nucleus is mainly due to quasicrossings of adiabatic terms
associated with the reaction channels. Let us specify the
right-centre states by the parabolic quantum numbers $n'_1$,
$n'_2$\,, and the principle quantum number $n'=n'_1+n'_2+m+1$. It
is known~\cite{Ponom2} that the quasicrosssings occur for terms
with the same quantum numbers $m$ and $n_1$. As the state $\psi_0$
has the zeroth values of these numbers, let us confine ourselves
to right-centre states with $m=n'_1=0$. Their wavefunctions have
no nodes in the variable $\xi$, but differ in the number $n_\eta$
of nodes in the variable $\eta$. The states with $5\leq n'_2\leq
10$ will be of interest in the following discussion. According to
the relations between $n_\eta$ and $n'_2$ presented in the
treatise~\cite{Ponom2}, for these states $n_\eta=n'_2$ at
$Z\,'\approx 11.1$\,. The dependences of the eigenvalues
${\varepsilon_j}$ of the two-centre problem~(\ref{2cent}) on the
interatomic distance $R$ are shown in Figure~\ref{eps_R} for the
%
%
%
%
%
%
states with $n'_2=5-8$\,; the subscript $j$ is now reduced to the
parabolic quantum number $n'_2$\,. There are quasicrossings at
$R\approx 8\,, 13\,,\mbox{ and } 21$. As $R$ is increased, the
value of $n'_2$ increases by unity at each quasicrossing. This
corresponds to the general rule~\cite{Ponom2} that the terms
involved into a quasicrossing differ in the number $n_\eta$ by
unity. There are three more quasicrossings not shown in
Figure~\ref{eps_R}. Two of them lie at $R\approx 37\mbox{ and }
85$. In passing through each of these quasicrossings in the
direction of growth of $R$, the number $n'_2$ increases by unity
and takes on the values of 9 and 10. Finally, the outermost
quasicrossing lies at $R\approx 878$. It involves the right-centre
state with $n'_2=10$ and the state $\psi_0$ for which the number
of nodes $n_\eta=11$.

Let us assume that muonic protium and neon are separated by a very
large distance $R$. The function $\psi_0$ corresponding to this
case is localized near the proton and practically identical there
to the atomic $1S$--state wavefunction. All their eleven nodes are
located near the neon nucleus where $\psi_0$ is exponentially
small. Such a situation survives to the outermost quasicrossing
with the right-centre state for which $n'_2=10$. After passing a
very narrow quasicrossing region, the muon charge distribution in
these states changes drastically. In the state $\psi_0$ the muon
charge cloud migrates to neon and becomes exponentially small near
the proton. In the state with $n'_2=10$ everything is opposite:
the charge flows to the proton and all the nodes of the
wavefunction prove in a region near neon where the wavefunction is
exponentially small. A similar situation is observed in passing
through the other long-distance quasicrossings which occur deep
under the potential barrier separating the Coulomb wells of the
two-centre problem. Therefore, it is valid to say that between
narrow quasicrossing regions the muon charge cloud is localized
near one of the Coulomb centres and, as $R$ is reduced, the
right-centre states with the number $n'_2$ decreasing in
successive unit steps describe muonic protium in the Coulomb field
of neon. In particular, in the interval $21<R<37$ this is the
state with $n'_2=8$. The fact that it corresponds to muonic
protium in the field of neon is confirmed by calculations of the
adiabatic potential which is equal to the sum of the eigenvalue
\mbox{$m_{\mu\rm H}\cdot\varepsilon_j (R)$} of the Hamiltonian
${\hat H}_\mu$ and the mean value of the Coulomb repulsion of the
nuclei. At $R\sim 30$ this potential agrees with the polarization
potential $U_p(R)$ with one percent accuracy. With further
decrease in $R$ the quasicrossings occur nearer and yet nearer to
the barrier top, the quasicrossing regions become broader, and the
statement on the muon localization near one of the nuclei loses
its meaning. For example, the quasicrossing at $R\approx 8$ occurs
near the barrier top.

According to the standard viewpoint, the muon transfer is due to
not too distant quasicrossings occurring near the barrier top. For
example, the muon transfer to carbon and oxygen results from
quasicrossings at $R=7-9$~\cite{Gersh}, the transfer to fluorine
is due to the ones at $R\approx 12$~\cite{Holz}. In
ref.~\cite{Sayas} the transfer to neon was attributed to
quasicrossings at $R\approx 26$. Therefore, in the present work
only the four right-centre states with the quantum numbers
$m=n'_1=0$ and $n'_2=5-8$ were taken into account in the expansion
of the three-body wavefunction (Fig.~\ref{eps_R}). In this case
there is a set of four coupled radial equations in the region
lying on the left of the quasicrossing at $R\approx 21$. On the
right of this quasicrossing, the state with $n'_2=8$ describes
muonic protium in the field of neon. The matrix elements coupling
the equation for the radial function of this state with the other
equations fall exponentially in increasing $R$. Therefore, at
$R\sim~30$ this equation is separated from the others and
corresponds to the entrance channel of the transfer reaction. As
it was already mentioned, at this value of $R$ the adiabatic
potential in this equation agrees with the polarization potential
with one percent accuracy. At $R>30$ the equation~(\ref{xi0}) with
the polarization potential $U_p(R)$ was used for the description
of the entrance channel, i.e. all the deep subbarrier
quasicrossings lying at $R\geq 37$ were fully ignored. In this
case the transfer channel is described by the three radial
equations for the states with $n'_2=5-7$.

The above consideration related to the muon transfer to a bare
nucleus $Z$. Actually, muonic hydrogen collides with an atom or a
molecule which have an electron shell. The energy gain in the
transfer reaction is a few keV, and it is more than enough for an
electron excitation. An analysis of the dynamics of the electron
shell during the collision is a complicated problem including a
construction of electron terms in the Coulomb field of the
three-body system $\mu{\rm H}Z$ and an examination of transitions
between them. The simplest approximation is to ignore any
excitations and to assume that the electron shell remains in its
ground state during the collision. In this case the role of the
electron shell is reduced to the screening of the Coulomb
interaction of heavy particles in the reaction channels. It is
natural to expect that because of small collision energies the
screening is most significant in the entrance channel. In the
present work the screening was taken into account in the
equation~(\ref{xi0}) which descibes the entrance channel at
$R>30$. Instead of the polarization potential $U_p(R)$, a new
potential $U_e(R)$ was used in this equation. The method of its
construction was suggested in ref.~\cite{Krav}. For the
interaction of muonic hydrogen with a noble gas atom at collision
energies below the lowest excitation energy of the electron shell
(16.6~eV for neon), the potential $U_e(R)$ is:
\begin{equation}
\label{Ue}
U_e(R)=U_s(R)+U_f(R)+U_w(R)\,.
\end{equation}
The first term is the screened polarization potential:
\begin{equation}
\label{Us}
U_s(R)=-\frac{\beta Z_a^{\,2}(R)}{2R^4}\,,
\quad
Z_a(R)=Z-Z_e(R)\,.
\end{equation}
$Z_e(R)$ is the absolute value of the electron charge inside the
sphere of the radius $R$ centered at the nucleus $Z$,\, $Z_a(R)$
is the total atomic charge in this sphere. The term $U_f(R)$ may
be treated as a contact interaction of muoinic hydrogen with the
electron shell:
\begin{equation}
\label{Uf}
U_f(R)=\frac{2\pi}{3}\,<r_{\mu\rm H}^2>\rho_e(R)\,.
\end{equation}
$<r_{\mu\rm H}^2>$ is the mean-square charge radius of muonic
hydrogen in the $1S$--state. It is calculated with respect to the
centre of mass of $\mu\rm H$:
\begin{equation}
\label{r2}
<r_{\mu\rm H}^2>=-\frac{3}{m_{\mu\rm H}}\left(1-\frac{1}{M_{\rm
H}}\right)\,.
\end{equation}
This quantity is negative because it is mainly contributed by the
negatively charged muon. The function $\rho_e(R)$ is the absolute
value of the electron density at the distance $R$ from the nucleus
$Z$. It is normalized as follows:
\begin{equation}
\label{roe}
4\pi\int\limits_0^\infty\rho_e(R)\,R^2 dR=Z\,.
\end{equation}
The charge $Z_e(R)$ and the density $\rho_e(R)$ were calculated
with analytical one-electron wavefunctions obtained within the
Roothan-Hartree-Fock method~\cite{Clem}. Both the potentials
$U_s(R)$ and  $U_f(R)$ are attractive and fall exponentially with
increasing $R$. As $U_s(R)$ is proportional to the square of
$Z_a(R)$ and, in addition, to $R^{-4}$, it falls faster. As a
result, this potential is significant at distances $R$ less than
the electron Bohr radius ($\approx 200$~m.a.u.).  For example, at
$R=30$ the potential $U_s\approx -1.7~\rm\mbox{eV}$ and it exceeds
$U_f$ by an order of magnitude. As the electron $K$--shell of neon
is similar in size, the screening effect on the potential $U_s$ is
already noticeable: the charge $Z_a\approx 8.9$\,. At $R\approx
100$ the potentials $U_s$ and $U_f$ become equal to each other and
their sum is about $-0.02~\rm\mbox{eV}$, i.e. it is of the order
of thermal energies at room temperature. At $R\approx 200$ the
term $U_s$ is already about 5\% of $U_f$. The latter is equal to
$-0.002~\rm\mbox{eV}$.

The last term $U_w(R)$ in the formula~(\ref{Ue}) appears in the
second order of the perturbation theory with respect to the
Coulomb interaction of atomic electrons with muonic hydrogen. It
corresponds to the van der Waals attraction at large $R$. An
accurate calculaton of this potential is not a simple matter
because it involves the summation over intermediate states of both
the electron shell and muonic hydrogen. The asymptotic expansion
of this potential in powers of $R^{-1}$ was considered in
ref.~\cite{Roman}. Its leading term has the standard form
$(-C/R^6)$. The constant $C$ was estimated in the completeness
approximation: $C=1.90\times 10^{+6}$~m.a.u. In this case the
potential $U_e(R)$ was written in the simplest form which provides
the correct asymptotic behaviour at large $R$:
\begin{equation}
\label{Uew}
U_e(R)=\left\{
\begin{array}{ll}
U_s(R)+U_f(R)&,\,R<R_w\,;\\
-C/R^6&,\,R\geq R_w\,.\\
\end{array}
\right.
\end{equation}
$R_w$ is the interatomic separation at which the sum
$U_s(R)+U_f(R)$ becomes equal to the van der Waals potential
$(-C/R^6)$. For neon $R_w\approx 1070$~m.a.u. ($\approx
2\,.7$~\AA) and the potential at this distance is very small:
$U_e(R_w)\approx 7\times 10^{-9}$~eV. Therefore, the van der Waals
tail of the potential is insignificant at the considered collision
energies $E_c\ge 10^{-4}$~eV, and it is possible to set
$U_e(R)=U_s(R)+U_f(R)$ for all $R$.

In order to clarify the role of the electron screening, the
calculations were made for the same cases A, B, and C as in
ref.~\cite{Roman} (Sect.~\ref{Intro}). These cases differ in the
potential in the equation~(\ref{xi0}) asymptotically describing
the entrance channel. The unscreened polarization potential
$U_p(R)$ was used in the case A. This corresponds to the muon
transfer to the bare neon nucleus. The screened potential $U_s(R)$
was substituted for $U_p(R)$ in the case B, and the potential
$U_e(R)=U_s(R)+U_f(R)$ was employed in the case C. In all these
cases, for nonzero angular momenta $J$ there is a barrier in the
effective potential appearing in the equation~(\ref{xi0}). The
position $R_b$ of the barrier top and its height $U_b$ are given
in Table~\ref{tab_bar} for $J\leq 3$. For these values of $J$ the
%
%
%
%
\begin{table}
\caption{The position $R_b$ of the top of the effective-potential
barrier and its height $U_b$ for some angular momenta $J$ in the
cases A--C. In each box the upper value is $R_b$ in m.a.u., the
lower value is $U_b$ in eV.}
\bigskip
\begin{tabular}{|c|l|l|l|}
\hline
$\hspace{4mm}J\hspace{4mm}$&\hspace{6.5mm}A   &\hspace{6.5mm}B   &\hspace{6.5mm}C   \\
\hline
$1$                        &\hspace{3mm}76.1  &\hspace{3mm}62.4  &\hspace{3mm}65.6  \\
                           &\hspace{5mm}0.0519&\hspace{5mm}0.0841&\hspace{5mm}0.0672\\
\hline
$2$                        &\hspace{3mm}44.0  &\hspace{3mm}39.0  &\hspace{3mm}40.5  \\
                           &\hspace{5mm}0.467 &\hspace{5mm}0.641 &\hspace{5mm}0.577 \\
\hline
$3$                        &\hspace{3mm}31.1  &\hspace{3mm}28.9  &\hspace{3mm}29.5  \\
                           &\hspace{5mm}1.87  &\hspace{5mm}2.32  &\hspace{5mm}2.17  \\
\hline
\end{tabular}
\label{tab_bar}
\end{table}
%
%
barrier top lies in the region $R>30$ in which the entrance
channel is described by the equation~(\ref{xi0}). It is
interesting that the weakening of the attraction caused by the
electron screening leads to the barrier top being shifted to lower
$R$ and its height increasing. This fact was pointed out in
ref.~\cite{Ital}. At low collision energies $E_c\ll U_b$\,, the
barrier prevents the penetration of the corresponding partial wave
into the interaction region. As a result, the contribution of this
wave to the transfer cross-section is small. As the collision
energy goes up to the barrier top, the partial transfer
cross-section increases and becomes comparable to contributions of
waves with lower angular momenta. Moreover, quasi-steady states
may exist under the barrier. The transfer cross-section has a
resonance peak in a vicinity of such a state. In ref.~\cite{Roman}
one quasi-steady state was found in the $D$--wave at collision
energies of 0.3--0.5~eV.
%
%
\section{Results of the calculations and conclusions}
\label{Res}
\begin{figure}
\includegraphics{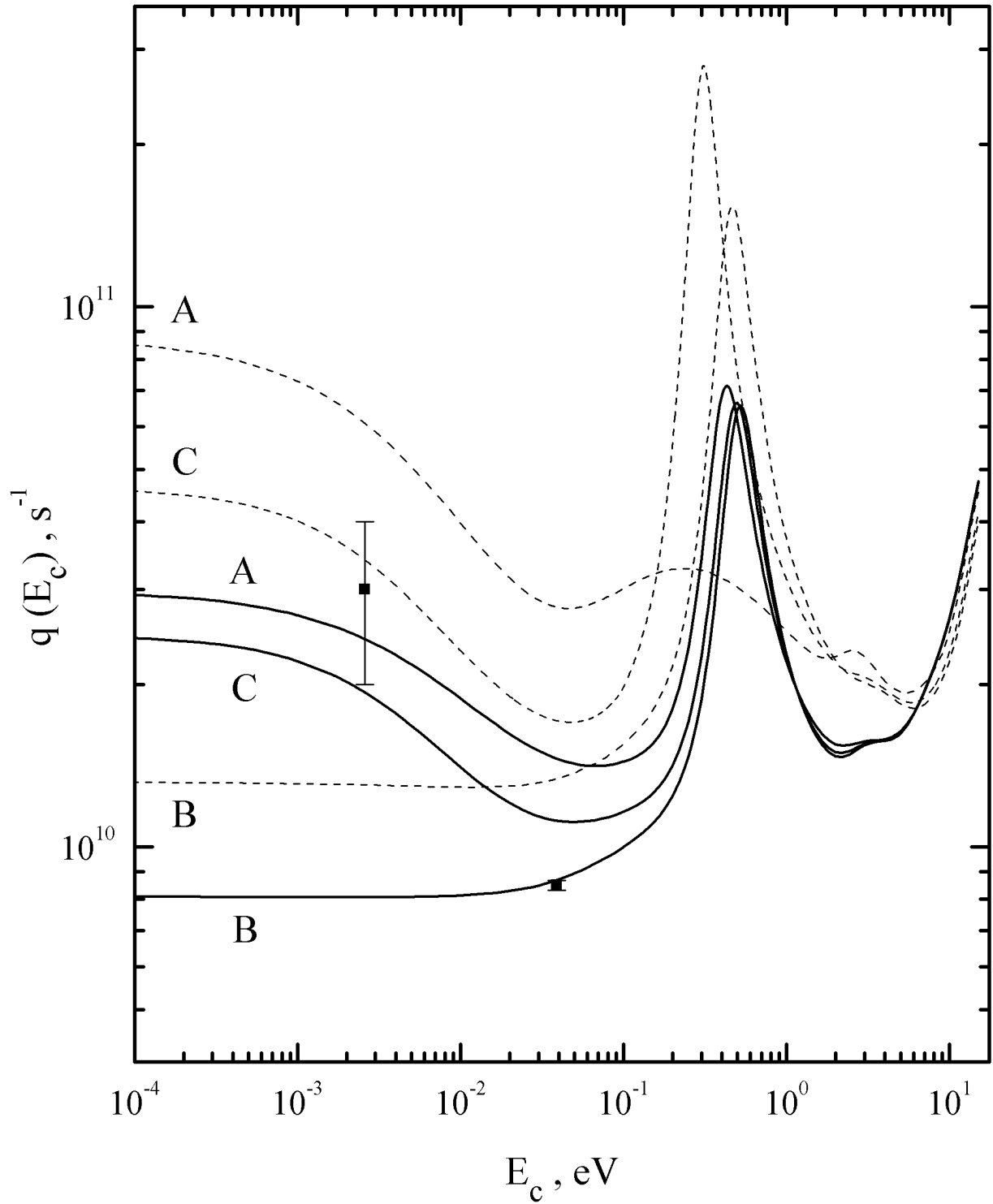}
\caption{The muon transfer rate $q(E_c)$ vs. the collision energy
$E_c$. The rate is reduced to the atomic density of liquid
hydrogen. The solid curves are the results of the present work,
the dashed curves are the results obtained in ref.~\cite{Roman}.
The curves are marked with the letters A, B, and C in accordance
with the three ways of taking into account the electron screening.
The experimental values of the transfer rate (Table~\ref{tab_res})
are attributed to the mean thermal energies $(3/2)kT$ at $T=20$
and 300~K.}
\label{rate_E}
\end{figure}

%
%
The energy dependences of the transfer rate $q(E_c)$ are shown in
Figure~\ref{rate_E}. They are similar to those found in
ref.~\cite{Roman}. At low collision energies, the $S$--wave makes
the main contribution, the cross-section is proportional to
$v^{-1}$, and the transfer rate is nearly constant. Its values
depend on the way of taking into account the electron screening,
but in a less degree than in ref.~\cite{Roman}. With a rise in the
collision energy, the transfer rate obtained in the cases A and C
decreases, while in the case B it remains almost constant. The
contribution of the $P$--wave becomes significant at energies
corresponding to room temperature,  and the transfer rate begins
to go up gradually. As a result, in the cases A and C there is a
broad minimum covering the region of thermal energies. This
corresponds qualitatively to the strong suppression of the
transfer reaction at room temperature. On the whole, at
$E_c<0.1$~eV the curves of $q(E_c)$ found in the present work are
obtained by moving the curves of ref.~\cite{Roman} down. With
further increase in the collision energy, the $D$--wave begins to
play a dominant role. In all the cases A--C, there is a resonance
peak on the curve $q(E_c)$ at $E_c\approx 0.5$~eV. It is due to
the already mentioned quasi-steady state in the $D$--wave.
Parameters of this peak, such as its position, width, and height,
depend slightly on the electron screening. It is interesting that
in ref.~\cite{Roman} the resonance peak was obtained only in the
cases B and C, and its parameters were more sensitive to the
electron screening. Moreover, compared to the present work the
peak was noticeably higher and wider. The effect of the $D$--wave
resonance is manifested up to the energy $E_c\approx 2$~eV. Then
the transfer rate passes through one more minimum, and at
$E_c>5$~eV it begins to go up again due to the contribution of
waves with $J\geq 3$. The electron screening is already
insignificant in this region, and the curves obtained in the cases
A--C are practically identical. It should be noted that the rapid
increase of the muon transfer rate at $E_c>0.2$~eV is of interest
in connection with the question of a measurement of the hyperfine
splitting of the $1S$--state of muonic protium~\cite{Fr1}.

\begin{figure}
\includegraphics{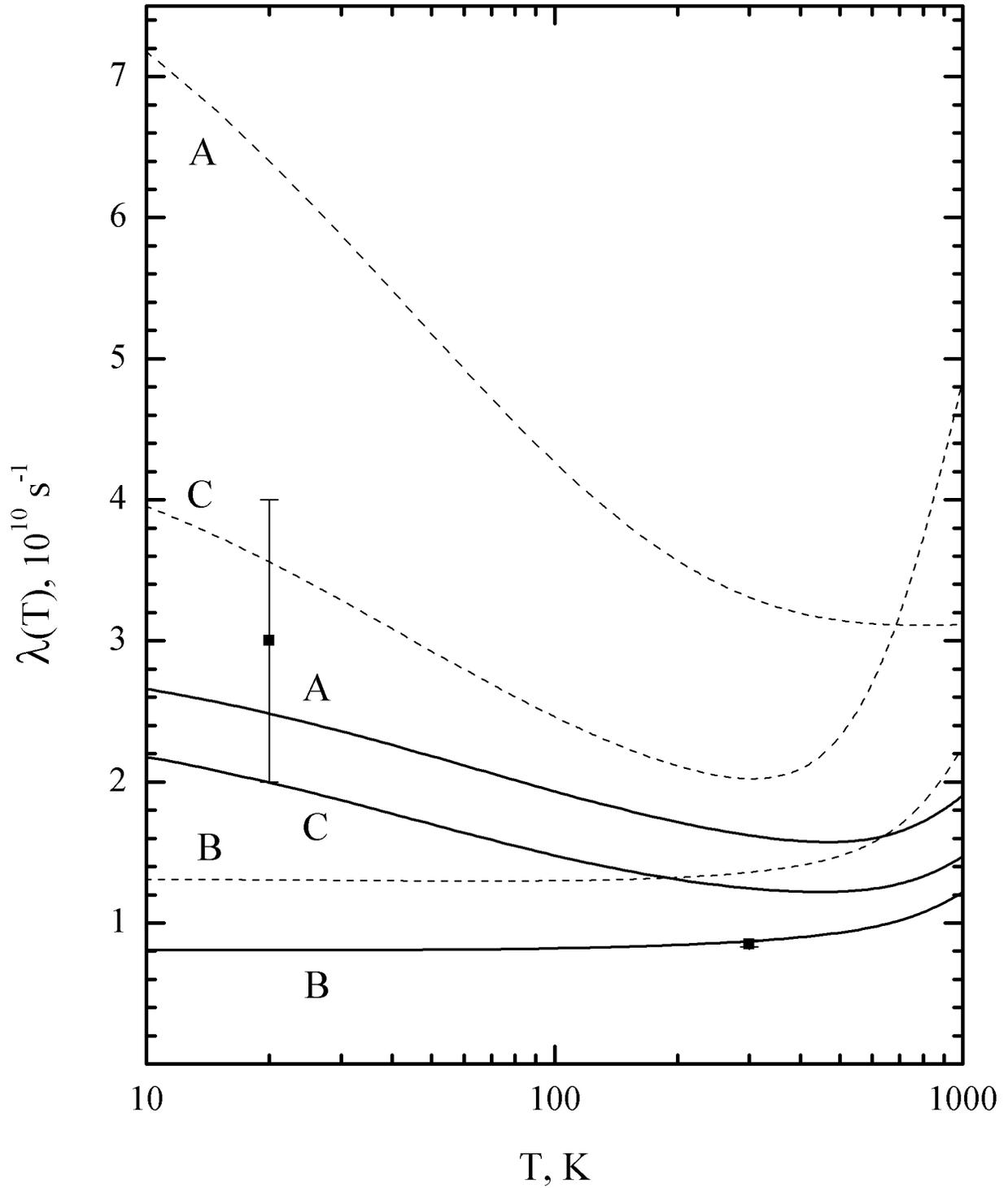}
\caption{The rate $\lambda(T)$ of the muon transfer from
thermalized $\mu p$ atoms vs. the temperature $T$. The rate is
reduced to the atomic density of liquid hydrogen. The notations of
the curves are identical to those used in Figure~\ref{rate_E}. The
experimental values correspond to the temperatures $T=20$ and
300~K (Table~\ref{tab_res}).}
\label{rate_T}
\end{figure}

The temperature dependences of the rate $\lambda(T)$ of the muon
transfer from thermalized $\mu p$ atoms are shown in
Figure~\ref{rate_T}. They were obtained by averaging $q(E_c)$ over
the Maxwellian distribution of relative velocities in the entrance
channel. The values of $\lambda(T)$ at the temperatures of 20~K
and 300~K are given in the last column of Table~\ref{tab_res}. The
curves obtained in the present work lie below the corresponding
curves of ref.~\cite{Roman}. The curve found in the most realistic
case C correctly reproduces the tendency to decreasing the
transfer rate with increasing the temperature in the interval
20--300~K\,. At $T=20$~K it passes through the lowest point of the
interval of experimental values. At $T=300$~K the calculated value
of $\lambda(T)$ exceeds the experimental value by a factor of
1.5\,. The curve obtained in the case A lies at greater values of
the transfer rate, and it is nearly parallel to the curve C. The
value of $\lambda(T)$ found in the case B for room temperature
agrees very well with the experiment (the accuracy is about 2~\%),
but at lower temperatures the transfer rate is nearly constant. At
high temperatures the values of $\lambda(T)$ obtained in the case
C go up more slowly than in ref.~\cite{Roman}. This is due to the
shift of the resonance peak on the curve of $q(E_c)$ to greater
energies and the decrease of its width and height. On the whole,
the results obtained in the present work agree with the
experimental data better than the results of ref.~\cite{Roman}.
The effect of the electron screening proves to be somewhat less,
but still noticeable.

This work was supported by the Grant NS--215.2012.2 from the
Ministry of Education and Science of the Russian Federation.
%
%
%
%
%
%
%
%
%
%
%
%

%
%
\end {document}